\documentclass[preprint]{aastex}
\usepackage{mkfig}

\begin{document}
 
\title{\bf Elemental Abundances and Ionization States within the Local
  Interstellar Cloud Derived from HST and FUSE Observations of the Capella
  Line of Sight\altaffilmark{1}}

\author{Brian E. Wood\altaffilmark{2}, Seth Redfield\altaffilmark{2},
  Jeffrey L. Linsky,\altaffilmark{2} and M. S. Sahu\altaffilmark{3}}

\altaffiltext{1}{Based on observations made with the NASA-CNES-CSA Far
  Ultraviolet Spectroscopic Explorer.  FUSE is operated for NASA by the
  Johns Hopkins University under NASA contract NAS5-32985.  Also based
  on observations with the NASA/ESA Hubble Space Telescope, obtained from
  the Data Archive at the Space Telescope Science Institute, which is
  operated by the Association of Universities for Research in Astronomy,
  Inc., under NASA contract NAS5-26555.}
\altaffiltext{2}{JILA, University of Colorado and NIST, Boulder, CO
  80309-0440; woodb@origins.colorado.edu, jlinsky@jila.colorado.edu.}
\altaffiltext{3}{Goddard Space Flight Center, Code 681, Greenbelt, MD
  20771; msahu@panke.gsfc.nasa.gov}

\begin{abstract}

     We use ultraviolet spectra of Capella from the {\em Hubble Space
Telescope} (HST) and {\em Far Ultraviolet Spectroscopic Explorer} (FUSE)
satellites to study interstellar absorption lines from the Local
Interstellar Cloud (LIC).  Measurements of these lines are used to
empirically determine the ionization states of carbon, nitrogen, and
silicon in the LIC, for comparison with the predictions of theoretical
photoionization models.  We find that the observed ionization
states are consistent with previously published photoionization
predictions.  Total abundances are determined for the elements
mentioned above, and others, for comparison with solar abundances.
Magnesium, aluminum, silicon, and iron are all depleted by at least a
factor of 10 toward Capella.  The abundances of carbon, nitrogen, and
oxygen are essentially solar, although the error bars are large enough
to also allow depletions of about a factor of 2 for these elements.

\end{abstract}


\section{INTRODUCTION}

     The Sun is located inside a warm, partially ionized interstellar
cloud called the Local Interstellar Cloud (LIC).  Ultraviolet spectra
containing LIC absorption lines have been used to study
the properties of the LIC.  Its average temperature appears to be about
$T=8000\pm 1000$~K, although there is some evidence for spatial
variation within the LIC \citep{jll95,ard97,np97,bew98,bew02a}.
The average H~I density is $n({\rm H~I})\approx 0.1$ cm$^{-3}$ \citep{jll00}.
Hydrogen must be roughly half-ionized since the average electron density
is $n_{e}=0.11^{+0.12}_{-0.06}$ cm$^{-3}$ \citep{bew97,jbh99}.
Estimates of $n({\rm H~I})$ from observations of
interstellar atoms passing through the heliosphere suggest somewhat higher
H~I densities of $n({\rm H~I})\approx 0.2$ cm$^{-3}$ \citep{eq94,vvi99},
possibly suggesting the presence of
density and/or hydrogen ionization state variations within the LIC.
Ionization state variations are to be expected since different parts of
the LIC will be shielded from photoionization sources to different extents
\citep{fcb88,kpc90,jds02}.  Measurements of interstellar material flowing
through the heliosphere and local interstellar medium (LISM) absorption
line studies all suggest that in a heliocentric rest frame the LIC appears
to be flowing towards Galactic coordinates $l=186.1^{\circ}$ and
$b=-16.4^{\circ}$, with a speed of about 25.7 km~s$^{-1}$
\citep{mw93,mw96,rl92,rl95,eq00}.

     Probably the best line of sight for studying absorption by gas
in the LIC is that toward Capella, which is a spectroscopic binary
system (G8~III+G1~III) located 12.9~pc away, with Galactic coordinates
$l=162.6^{\circ}$ and $b=+4.6^{\circ}$.  There are several reasons why
this line of sight is ideal.  First of all, there is only one interstellar
velocity component observed for this short line of sight, that
of the LIC.  Short lines of sight are preferable for studying the LIC
to avoid a complicated, multi-component ISM structure that is often
difficult to resolve into individual components.  Secondly, although
cool stars are not as bright in the UV as hot stars, Capella
is the brightest cool star in the sky in the ultraviolet, providing
sufficient background in the continuum and bright emission lines to observe
numerous LIC absorption lines.  There are few nearby hot stars (including
hot white dwarfs) that can be observed for LIC studies, which means one
must generally observe cool stars like Capella.  The UV emission lines
from the Capella stars are quite broad, making it easy to distinguish
narrow ISM lines located within the broad emission profiles, which serve
as the continuum for measuring the ISM lines.

     Another advantage of the Capella line of sight is that LIC
column densities are particularly high in this direction, allowing
us to detect weak absorption lines that would be undetectable in other
directions.  This is illustrated by Figure~1, a schematic picture of the
structure of the LISM in the Galactic plane, showing
that the Capella line of sight passes through the entire length of the LIC.
The LIC outline in Figure~1 is from the model of \citet{sr00}.
The shape of the nearby ``G cloud'' is a crude estimate from \citet{bew00},
and there is some question as to whether this
cloud is truly distinct from the LIC \citep*[see][]{bew02a}.  Also shown
is an equally crude estimate of the shape of a cloud detected toward Sirius
(dotted line), which might be the same cloud as the ``Hyades Cloud''
identified by \citet{sr01} in the direction of the Hyades
Cluster.  The lines of sight to the stars shown in Figure~1 have all
been studied by the {\em Hubble Space Telescope} (HST) to better understand
the properties of the LISM.  Since all the stars are $<10^{\circ}$ from the
Galactic plane, distortions due to the projection effects in showing the
three-dimensional LISM as a two-dimensional figure are not severe.  The
direction toward the B2~II star $\epsilon$~CMa is indicated in
Figure~1.  This star is important because it is the dominant stellar
source of ionizing photons irradiating the LIC \citep{jvv95}.

     Several studies of the Capella line of sight have been
published using data from the Goddard High Resolution Spectrograph (GHRS)
formerly aboard HST.  \citet{jll93,jll95} analyzed LIC
absorption lines of H~I, D~I, Mg~II, and Fe~II toward Capella, with the
primary goal being to measure the D/H ratio within the LIC.  \citet{avm98}
also analyzed the HST/GHRS data and found close
agreement with the results of \citet{jll93,jll95}, although
\citet{avm02} claim that the analysis of H~I should have
larger uncertainties due to the possible existence of undetected hot
H~I absorption components.  \citet{bew97} measured the electron
density toward Capella using observations of C~II $\lambda$1334.5 and
C~II$^{*}$ $\lambda$1335.7.

     One problem with the GHRS instrument is that its one dimensional
detector could only observe a relatively narrow wavelength region for each
exposure.  However, in 1997 the GHRS was replaced with the Space Telescope
Imaging Spectrograph (STIS).  The STIS instrument has capabilities similar
to the GHRS in terms of spectral resolution and sensitivity, but it has a
two-dimensional detector providing much broader wavelength coverage for
each observation.  On 1999 September 12,
HST/STIS observed the $1170-1710$~\AA\ spectrum of Capella using the E140M
grating.  This observation provides access to many LIC absorption lines
unavailable in the GHRS data set.  The {\em Far Ultraviolet Spectroscopic
Explorer} (FUSE) has also observed Capella recently.  The FUSE satellite
obtains spectra at wavelengths shorter than those accessible to HST, from
$905-1187$~\AA, allowing access to still more ISM absorption features.

     In this paper, we use the new FUSE and HST/STIS data to provide a
complete analysis of UV absorption lines detected towards Capella.
Our primary goal is to measure column densities for as many atomic species
as possible to establish the ionization states and abundances of various
elements within the LIC.

\section{OBSERVATIONS}

     Table~1 lists the HST/STIS and FUSE observations of Capella analyzed
in this paper.  The STIS instrument is described in detail by \citet{rak98}
and \cite{bw98}.  The stellar Fe~XXI $\lambda$1354.1 emission line in the
STIS spectrum was analyzed by \citet{oj02}, and the stellar emission lines
in the FUSE data were studied by \citet{pry01}.  Here we are obviously
interested in the LIC absorption lines present in these data.  The HST/STIS
data obtained on 1999 September 12 were processed using the IDL-based
CALSTIS software package \citep{dl99}.  This processing includes an
accurate wavelength calibration using an exposure of the onboard wavelength
calibration lamp obtained contemporaneous with the Capella exposures, and
a correction for scattered light within the spectrograph.

     On 2000 November 5, FUSE observed Capella through the
$30^{\prime\prime}\times 30^{\prime\prime}$ low resolution (LWRS) aperture.
The observation consisted of 10 separate exposures adding up to a total
exposure time of 14,182~s.  A nearly identical 10-exposure observation of
Capella was obtained two days later on 2000 November 7 with
a total exposure time of 12,332~s.  A third observation of Capella was
made through the $4^{\prime\prime}\times 20^{\prime\prime}$ medium
resolution (MDRS) aperture on 2001 January 11,
consisting of 28 separate exposures totaling 21,359~s.  These data were
reduced with version 1.8.7 of the CALFUSE pipeline software, although for
the H~I Ly$\beta$ spectral region, we found it necessary to use spectra
processed using the more recent CALFUSE 2.0.5 version (see \S3.4).

     Given Capella's orbital period of 104 days, the two LWRS observations
were made close enough in time (and therefore orbital phase) that we can
coadd the two data sets to produce
a single LWRS spectrum.  However, the MDRS data must be kept separate,
because this data set was obtained with the Capella binary at orbital
phase $\phi=0.48$ (near opposition), while the LWRS data were obtained at
$\phi=0.84$ (closer to quadrature).  Thus, the centroids of the emission
lines from the two Capella stars are different for the two data sets,
changing the composite line profiles.  The spectra must therefore be
analyzed separately.  Another reason for keeping the data separate involves
airglow emission.  The large size of the LWRS aperture means that H~I and
O~I airglow emission features from the Earth's geocorona are very prominent
in the LWRS spectrum, whereas the airglow features are greatly suppressed
in the MDRS data due to the much smaller size of the MDRS aperture.
Because of the reduced airglow, there are H~I, D~I,
and O~I absorption features visible in the MDRS data that are largely or
completely obscured in the LWRS data.  The existence of the separate MDRS
data set therefore allows us to analyze absorption lines that are not
visible in the LWRS data.

     In order to fully cover its 905--1187~\AA\ spectral range, FUSE has a
multi-channel design --- two channels (LiF1 and LiF2) use Al+LiF coatings,
two channels (SiC1 and SiC2) use SiC coatings, and there are two different
detectors (A and B).  For a full description of the instrument, see
\citet{hwm00}.  With this design FUSE acquires spectra in 8 segments
(LiF1A, LiF1B, LiF2A, LiF2B, SiC1A, SiC1B, SiC2A, and SiC2B) covering
different, overlapping wavelength ranges.  We cross-correlated and coadded
the individual FUSE exposures to create a single LWRS and MDRS spectrum for
each segment.  We decided not to coadd the individual segments to ensure
that such an operation would not degrade the spectral resolution.

     Since there is no wavelength calibration lamp onboard FUSE, the
wavelength scale produced by the FUSE data reduction pipeline is very
uncertain.  Fortunately, there are two strong, narrow interstellar
absorption lines in the Capella spectrum that can be used to fix the
wavelength scale for all of the segments except LiF2A and LiF1B
\citep[see][]{sr02}.  These
lines are C~III $\lambda$977.0 and C~II $\lambda$1036.3.  As mentioned
in \S 1, the Capella line of sight has only one interstellar velocity
component, which is identified with the LIC and centered at a heliocentric
velocity of $+22$ km~s$^{-1}$ \citep{jll93,jll95}.  By shifting
each spectral segment so that the ISM absorption is centered at this
velocity, we can calibrate the wavelength scales to within
$\sim 5$ km~s$^{-1}$.

\section{FITTING THE LIC ABSORPTION LINES}

\subsection{Carbon, Nitrogen, and Silicon}

     Our goal is to measure LIC column densities for as many atomic
species as possible using the full HST and FUSE data sets.  We searched
the spectra for absorption lines using the \citet{dcm91} line list.
Our initial attention focused on lines of carbon, nitrogen, and silicon,
because for the lowest three ionization states of these elements we
can either measure a column density from detected absorption lines
or measure a meaningful upper limit from undetected lines.
Figure~2 shows the detected and undetected lines that we use to
measure these column densities.
Note that there are 3 N~I and 3 Si~II lines used to measure the N~I and
Si~II columns.  We use the undetected C~I $\lambda$1656.9,
N~III $\lambda$989.8, Si~I $\lambda$1562.0, and Si~III $\lambda$1206.5
lines to estimate upper limits for those species.  The C~II $\lambda$1334.5
line in Figure~2 is not from our STIS data, but is from a GHRS spectrum
that has higher spectral resolution than the STIS data analyzed here.
Since the GHRS C~II line was previously analyzed by \citet{bew97},
it is not analyzed again here, but we reproduce the C~II profile and fit
in Figure~2 for the sake of completeness.

     Fits to the detected LIC absorption lines are shown in Figure~2.
Polynomial fits to the wavelength regions surrounding the absorption
lines are used to estimate the continuum above each line.  Best fits to
the lines are determined by a $\chi^{2}$ minimization routine.
Dotted lines show the fits before convolution with the instrumental
line spread function (LSF), and thick solid lines show the fits after
convolution with the LSF, which fit the data.  For the STIS data, we
use the appropriate LSF from \citet{kcs99}.  The FUSE LSF is
unfortunately not very well known.  Furthermore, it is wavelength
dependent and quite possibly changes from one observation to the next
\citep[see, e.g.,][]{jwk02}.  We use the average FUSE LSF derived by
\citet{bew02b} from the analysis of white dwarf spectra, recognizing
that this profile will only apply to our data in an approximate sense.
However, since we typically {\em assume} Doppler broadening
parameters rather than try to derive them from the fits (see below), we
do not believe that inaccuracies in the LSF will result in significant
errors in our results.  The three N~I lines in Figure~2 are fitted
simultaneously to measure the N~I column density, as are the three Si~II
lines to derive the Si~II column.

     None of the detected absorption lines are fully
resolved in either the STIS or the FUSE spectra.  Based on the analysis of
higher resolution GHRS observations of fully resolved H~I, D~I, Mg~II, and
Fe~II lines, \citet{jll93,jll95} measured a temperature and
nonthermal velocity for the LIC toward Capella of $T=7000\pm 900$~K and
$\xi=1.6\pm 0.6$ km~s$^{-1}$, respectively.  (The errors include both
random and systematic errors combined linearly).
We can use these measurements to constrain the Doppler broadening
parameters ($b$) of our lines using the relation
\begin{equation}
b^{2}=0.0165\frac{T}{A}+ \xi^{2},
\end{equation}
where $A$ is the atomic weight of the element in question and $b$ is in
units of km~s$^{-1}$.  As an example, for N lines ($A=14$) the
Doppler parameter can be constrained to be in the range
$b=2.86-3.75$ km~s$^{-1}$.  By constraining the Doppler parameters in
this way, we can obtain more precise column density measurements from the
unresolved spectral lines.  For the N~I and Si~II lines, we tried relaxing
this constraint since we had several lines of differing strengths to work
with (see Figure~2), which collectively provide constraints for $b$ even
though the lines are not resolved.  We did not obtain results significantly
different from the fits in which the value of $b$ is constrained.

     Table~2 provides a summary of all column density measurements for LIC
absorption lines observed toward Capella.  Some of the measurements are
based on previous work, as indicated by the last column of the table.  The
column densities we measure here are also listed in the table.
The new column densities quoted in Table~2 are actually the product of two
independent analyses.  Two of us (B.~E.~W.\ and S.~R.) reduced and analyzed
the STIS and FUSE data separately.  Although the measurement techniques are
essentially identical, as described above, the continuum placement will be
different since it will depend on exactly how broad a region beyond the
lines is fitted with the polynomial and what order of polynomial is used.
The fact that the data are reduced independently in these analyses is also
significant, especially for the FUSE data because the manner in which
individual exposures are cross-correlated and coadded can lead to
differences in the reduced spectrum.  Thus, this exercise allows us to see
if two important potential sources of systematic error are seriously
affecting our results.

     The independent measurements proved to be reasonably consistent in
that the error bars of the measured column densities overlap nicely,
suggesting that the dominant source of uncertainty is not continuum
placement or data reduction, but is instead the
uncertainty in the Doppler parameters, which arises from the uncertainties
in $\xi$ and $T$ (see above).  The independent measurements are certainly
not identical, though, and the new column densities and uncertainties
reported in Table~2 are averages of the two results.  For example, for
Si~II the independent analyses suggest $\log N({\rm Si~II})=12.94\pm 0.11$
and $\log N({\rm Si~II})=13.00\pm 0.05$, so the compromise value reported
in Table~2 based on averaging is $\log N({\rm Si~II})=12.97\pm 0.08$.

     The lines in Figure~2 that come from FUSE spectra are
C~III $\lambda$977.0, N~II $\lambda$1084.0, and N~III $\lambda$989.8.  The
data shown in the figure are the LWRS data, but we also fit the MDRS data,
and our derived column densities are compromises between the two separate
fits.  In Table~2, we list the segment used in the analysis of the FUSE
lines, although we check all segments for consistency.  For the C~III and
N~III lines, we choose the SiC2A segment, as the SiC1B segment that also
contains these lines is noisier and has significantly lower spectral
resolution than SiC2A.  For N~II, we use the SiC2B segment.  Figure~3
compares the N~II line seen by SiC2B with the SiC1A segment that also
contains this line, for both the LWRS and MDRS data.  For some reason the
LIC absorption feature is not seen in the SiC1A MDRS data, but its
detection in the other three cases leads us to conclude that the detection
is valid and that the MDRS SiC1A spectrum is misleading.  It is known that
FUSE's spectral resolution can decrease near the ends of segments.
Thus, perhaps the resolution of SiC1A degrades enough at the location of
the N~II line to make detection impossible, at least for the MDRS spectrum,
thereby forcing us to base our measurements on the SiC2B data.

     The uncertainties reported in Table~2 should be considered to be
roughly 2$\sigma$ uncertainties, although the issue is not free from
ambiguity.  For the new measurements presented in this paper, we always
assume $b$ values based on the $T$ and $\xi$ measurements and their
uncertainties from \citet{jll95}.  The random error contributions
to the $T$ and $\xi$ uncertainties reported by \citet{jll95}
are 2$\sigma$, but it is difficult to know what confidence level to
assign to the systematic error contributions that they estimate.  By adding
the two linearly rather than in quadrature we hope we are being
conservative enough to still consider the final total errors in $T$ and
$\xi$ (i.e., $T=7000\pm 900$~K and $\xi=1.6\pm 0.6$ km~s$^{-1}$) to be
$\sim 2\sigma$, meaning the column density errors in this paper reported
in this paper should also be $\sim 2\sigma$.  The H~I column in Table~2 is
from \citet{jll95}, but we assign a larger uncertainty to this
quantity based on the results of \citet{avm02} concerning
possible systematic errors involved in the analysis.  As is often the case
for systematic errors, the exact confidence level to associate with this
uncertainty is unclear.

\subsection{Oxygen}

     The O~I lines available for analysis are
shown in Figure~4.  The O~I $\lambda$1302.2 line observed by STIS
has excellent S/N but is the most highly saturated of the three
lines.  The O~I $\lambda$988.7, $\lambda$988.8, and $\lambda$1039.2
lines observed by FUSE are covered by airglow emission in the LWRS
spectra, so we can only use the MDRS data.

     Table~2 lists the segments used for the FUSE lines.  The
$\lambda$1039.2 line in Figure~4 is from the LiF1A channel.  The SiC1A
and SiC2B segments that also contain this line are too noisy to be of any
use, simply due to the lower sensitivity of the SiC detectors.  The
absorption feature is also not clearly detected in the LiF2B channel, but
the reason for this is unclear.  Nevertheless, we still consider the
LiF1A detection solid, because the absorption {\em must} be
present in the line at about the level seen in Figure~4 based on the
strength of the O~I absorption in the other lines shown in the figure.

     We fit the O~I lines separately, except for the adjacent
$\lambda$988.7 and $\lambda$988.8 lines, constraining the Doppler
parameter as described in \S3.1.  The best fits are shown in Figure~4,
and the column densities are listed in Table~2.  The weighted mean of
these three values for $\log N({\rm O~I})$ is 15.02.  We cannot use a
standard deviation for our error bar, because the uncertainties of the
three measurements are correlated through the assumed range
of $b$ values.  Thus, we simply use the smallest of the three
uncertainties, 0.32 dex, as our error estimate, resulting in a
final value of $\log N({\rm O~I})=15.02\pm 0.32$.  This is consistent with
the $\log N({\rm O~I})=14.91$ estimate of \citet{jll95} based on
a lower quality HST/GHRS observation of the O~I $\lambda$1302.2 line.

\subsection{Other Metal Lines}

     There is only one other metal line that is detected in the data,
Al~II $\lambda$1670.8.  This line and our best fit to it is
shown in Figure~5, and the column density
[$\log N({\rm Al~II})=11.43\pm 0.08$] is listed in Table~2.  The
$20.3\pm 0.9$ km~s$^{-1}$ velocity of the Al~II line agrees reasonably
well with the previously measured LIC velocity toward Capella of
$22.0\pm 0.9$ km~s$^{-1}$ \citep{jll95}, which helps convince us
that this weak LIC line is real.
Besides the C, N, and Si lines discussed in \S3.1, we also list in Table~2
upper limits for column densities of a few other metal lines that are
potentially useful:  C~IV $\lambda$1548.2, S~II $\lambda$1259.5, and
Ar~I $\lambda$1048.2.

\subsection{Hydrogen and Deuterium}

     The H~I and D~I column densities toward Capella have previously been
measured using HST/GHRS observations of the Ly$\alpha$ line
\citep{jll93,jll95}.  The more recent FUSE data allow access to the higher
lines of the Lyman series, although we must confine our attention to the
MDRS data since the LWRS data are contaminated with strong airglow emission.
There is D~I absorption detected in the Ly$\beta$ line at 1025.7~\AA, and
initially we hoped to analyze the Ly$\beta$ line to test the results of
the HST Ly$\alpha$ analysis.  This proved impossible, however,
partly due to problems with the FUSE ``walk correction'' (see below).
Nevertheless, the D~I detection and the data reduction difficulties for
Ly$\beta$ might be instructive and of some interest to the reader, so we
now describe the data and our problems with reducing and analyzing it in
more detail.

     The solid line in Figure~6 shows the LiF1A Ly$\beta$ profile that
is produced by processing the FUSE MDRS data through CALFUSE 1.8.7.
Vertical lines in the figure show where we expect the centroids of the D~I
and H~I absorption to be located.  There is an absorption feature at about
1025.6~\AA\ that looks like it could be D~I absorption, but it is
redshifted from where it should be by $\sim 0.1$~\AA\
($\sim 25$ km~s$^{-1}$), while the interstellar H~I absorption feature
is roughly centered correctly.  The problem can be traced to a detector
effect where the recorded position of each photon event is a slight
function of the pulse height of the event, which we
now describe in more detail.

     Each incident photon creates a cascade of electrons at the back of the
FUSE microchannel plate detectors.  When these electrons are read out by
the detector, the position of the event is recorded by the detector
electronics.  Unfortunately, the accuracy of this determination is degraded
for low pulse height events (i.e., events creating a weaker shower of
electrons).  This variation of position with pulse height is known as the
walk.  Furthermore, the FUSE detectors are gradually losing sensitivity
as they are continually exposed to more photons.  This ``gain sag'' means
a significant increase in low pulse height events and inaccuracies in the
wavelength positions of the detected photons \citep{djs00}.  This problem
is most severe at the location of Ly$\beta$, because the detector is
constantly exposed to substantial H~I Ly$\beta$ airglow emission.  This is
particularly true for LWRS data, but Figure~6 shows that it can be a
problem for MDRS data as well.  The gain sag that leads to the walk problem
can be mitigated by increasing the voltage of the FUSE detectors, thereby
increasing the gain.  So far, this has been done twice since launch.
Unfortunately, the first increase on 2001 January 24 happened just a couple
weeks {\em after} the MDRS observation of Capella, meaning the gain sag
had reached about its worst point at the time of the Capella observation
and has since been improved.

     Further calibration efforts by the FUSE instrument team have led to
a software correction for the walk problem, which is included in the more
recent version of the pipeline software, CALFUSE 2.0.5.  The Capella data
were obtained in TIME-TAG mode, where the individual pulse height
events associated with each detected photon are recorded and stored.  Thus,
we can reprocess the data with the walk correction included to try to more
accurately assign wavelengths for each event.  The dotted line in Figure~6
shows the LiF1A profile of Ly$\beta$ that results from data processing
through CALFUSE 2.0.5.  The improved data reduction shifts the D~I
absorption to roughly the correct location.  Note that it also broadens
the H~I absorption.

     In Figure~7, we show the Ly$\beta$ line from all four FUSE segments
that contain the line, after processing with CALFUSE 2.0.5.  Also shown are
estimates of the intrinsic stellar profile, and predictions for what the
D~I and H~I absorption should look like based on the H~I and D~I column
densities measured from Ly$\alpha$ \citep{jll95}.  There are some
subtle differences in the Ly$\beta$ profile as seen by the various FUSE
segments, particularly in the neighborhood of D~I.  These differences lead
to slight differences in the shapes of the stellar Ly$\beta$ profile
estimated for each segment.  The D~I absorption seems stronger and more
clearly detected in the noisier SiC segments.  While D~I in the CALFUSE
1.8.7 spectrum in Figure~6 was highly redshifted, it now seems to be
slightly blueshifted in the CALFUSE 2.0.5 spectra.  All these issues may
be indications of residual difficulties with the walk problem.  In any
case, we do not believe the FUSE Ly$\beta$ data to be of high enough
quality to improve upon the HST Ly$\alpha$ results, but Figure~7 shows
that the D~I and H~I Ly$\beta$ absorption is observed to be at least
roughly consistent with the HST Ly$\alpha$ analysis.

\section{THE IONIZATION STATES OF C, N, AND Si WITHIN THE LIC}

     The column density measurements and upper limits in Table~2 include
measurements for the three lowest ionization states of C, N, and Si.
Based on these measurements, we can compute the ionization fractions for
these elements within the LIC.  Because upper limits are involved (for C~I,
N~III, Si~I, and Si~III), we cannot quote a best value and an error for the
ionization fractions, so instead we compute ionization fraction ranges
allowed by the measurements.  These ranges are listed in Table~3 and are
also displayed graphically in Figure~8.  At least 95\% of carbon is in the
form of C~II.  Nitrogen is divided roughly equally between N~I and N~II,
although uncertainties are high due to the imprecise N~II measurement.
At least 90\% of silicon is Si~II.  Since the doubly ionized species
(C~III, N~III, and Si~III) have only trace abundances, we are confident
that higher ionization states are even more insignificant, and we note that
there is clearly no C~IV $\lambda$1548.2 or Si~IV $\lambda$1393.8
absorption toward Capella.

     In Figure~8, we compare the ionization fraction ranges allowed by
the Capella observations with the predictions of a photoionization model
from \citet[][hereafter SF02]{jds02}, in particular their favored
model 17.  In their models, SF02 estimate the ionizing extreme ultraviolet
(EUV) radiation field incident on the LIC based on the known stellar EUV
sources, dominated by $\epsilon$~CMa (B2~II), and estimates for the more
poorly known diffuse EUV background.  Radiative transfer calculations
are then used to compute ionization fractions at the
solar location within the LIC.  The comparison of these model predictions
with our measurements in Figure~8 is only valid in an approximate sense,
because our measurements are average values for the Capella line of
sight rather than for the LIC material in the immediate vicinity of the
Sun.  Nevertheless, despite this difference Figure~8 shows that the
observed ionization states of C, N, and Si toward Capella are
consistent with the predictions of SF02.

     The models of SF02 rely heavily on observations of LIC absorption
towards $\epsilon$~CMa \citep{cg95,cg01}.  \citet{cg01}
detect C~IV and Si~III absorption at the LIC velocity
toward $\epsilon$~CMa, in contrast to our low upper limits for these
species (see Table~2).  Numerically, \citet{cg01} find
${\rm C~IV/C~II}\approx 7\times 10^{-3}$ and ${\rm Si~III/Si~II}\approx 0.5$.
These results are inconsistent with our upper limits,
${\rm C~IV/C~II}< 2.5\times 10^{-3}$ and ${\rm Si~III/Si~II}< 0.01$.  The
Si~III discrepancy is particularly striking, and SF02 comment on the
difficulty of explaining any substantial amount of Si~III existing within
the LIC, although the low C~IV column could possibly be explained as
absorption from a conductive interface with the hot ISM.  However, since we
see no C~IV and Si~III absorption towards Capella it seems likely that the
absorption seen toward $\epsilon$~CMa is not really from the LIC, but is
instead from another component that happens to be at the LIC
velocity along the long line of sight to $\epsilon$~CMa.  This was also the
conclusion of \citet{gh99}, who noted that no Si~III
absorption is seen toward Sirius, a line of sight very near that of
$\epsilon$~CMa but much shorter (see Fig.~1).

     The agreement between the measured and predicted ionization states in
Figure~8 supports the view that photoionization alone can account
for the ionization state of the LIC.  However, SF02 presented a total of
25 different photoionization models for the LIC, and all 25 are consistent
with our observations, meaning our measurements are unfortunately not
precise enough to be be used to test the assumptions behind the individual
models.  The size of the error bars in Figure~8 also means the test
of the photoionization models is not as stringent as we would like, and in
any case it remains possible that elements other than C, N, and Si are
{\em not} in photoionization equilibrium.  \citet{ebj00} present another
argument in favor of photoionization equilibrium within the LISM in
general, pointing out that the low Ar~I/H~I ratios observed toward
several nearby white dwarfs are consistent with expectations for
photoionization, since Ar has a large photoionization cross section.
Unfortunately, we were not able to detect the Ar~I $\lambda$1048.2 line
toward Capella, and the upper limit in Table~2 is not very helpful.

     An alternative to the steady state photoionization model is the
possibility that the LIC is not in ionization equilibrium at all but was
collisionally ionized in the distant past ($\sim 10^{6}$ years ago) by a
passing shock wave, perhaps from a supernova, and ionization equilibrium
has still not been reestablished \citep{chl96}.  The high
ionization state of He within the LISM, as determined from {\em Extreme
Ultraviolet Explorer} (EUVE) measurements \citep{sv93,jbh95,jd95,tl96},
appears to support this view, since the known stellar EUV background
certainly cannot ionize He to the observed extent \citep{jvv98}.  Further
evidence for non-equilibrium conditions in the LISM is provided by
\citet{db01}, who has suggested that observations of X-ray and EUV emission
from the hot Local Bubble surrounding the LIC are more easily explained
by non-equilibrium models.

     For the LIC, the whole issue hinges on the nature of the diffuse
EUV background.  Unfortunately, the properties of this
emission are poorly known.  In their photoionization models, SF02 use
a measurement of the soft X-ray background from sounding rockets
\citep{dm83} and plasma models to infer the EUV background
from the hot plasma within the Local Bubble.  They also include another
diffuse component based on theoretical estimates of radiation from
evaporative boundaries of local clouds like the LIC.  With this assumed
background, SF02 can ionize He to the required degree, and their
predictions for C, N, and Si are consistent with our results, as mentioned
above.  However, upper limits on the diffuse EUV background provided by
the {\em Extreme Ultraviolet Explorer} (EUVE) satellite are lower than the
SF02 estimate by a factor of $5-10$ \citep{pj95}.
Fortunately, this issue should be resolved by the soon-to-be-launched
{\em Cosmic Hot Interstellar Plasma Spectrometer} (CHIPS) satellite, which
(unlike EUVE) is designed to measure diffuse EUV emission \citep{mh99}.

     Once CHIPS provides a more definitive measurement of the diffuse EUV
background, the whole issue of whether the LIC is in photoionization
equilibrium or is out of equilibrium entirely should be revisited.  For
now, we regard the comparison of theoretical and observed ionization
fractions in Figure~8 as providing some support for the accuracy of the
photoionization models of SF02, at least for the elements with which we
are concerned.

\section{ELEMENTAL ABUNDANCES WITHIN THE LIC}

     We now estimate total gas-phase abundances for the elements for
which we have column density measurements.  Absolute abundances must be
measured relative to the dominant element in the universe, hydrogen, so
we must first estimate the total hydrogen column density toward
Capella, including both H~I and H~II.  The H~I column has been measured
from HST/GHRS observations of Ly$\alpha$.  The H~I column that we assume
in Table~2 is from \citet{jll95}, but we assign a larger
uncertainty to this quantity based on the results of \citet{avm02}.
The H~II column density toward Capella cannot be measured directly.

     \citet{bew97} determined that the average electron density
within the LIC toward Capella lies in the range
$n_{e}=0.05-0.23$ cm$^{-3}$ based on the column density ratio of
C~II $\lambda$1334.5 and C~II$^{*}$ $\lambda$1335.7.  Most of the
electrons within the LIC will come from ionized hydrogen, but since
He is ionized to nearly the same degree as hydrogen in the LISM (see \S4),
roughly 10\% will come from He.  Thus, we
assume $n({\rm H~II})=n_{e}/1.1$.  This number density of H~II can be
converted to a column density only if we know the distance to the edge
of the LIC toward Capella, $d_{edge}$.  The highest average H~I densities
within the LIC toward the nearest stars are $n({\rm H~I})\sim 0.1$ cm$^{-3}$
\citep[see, e.g.,][]{jll00}, but estimates of $n({\rm H~I})$
within the LIC based on measurements of interstellar H~I within the
heliosphere have suggested that $n({\rm H~I})$ could be more like
0.2 cm$^{-3}$ \citep{eq94,vvi99}.
Assuming $n({\rm H~I})=0.1-0.2$ cm$^{-3}$, we find that
$d_{edge}=N({\rm H~I})/n({\rm H~I})=2.4-6.6$~pc based on the
$N({\rm H~I})$ value in Table~2.  We can then compute the H~II column
from $N({\rm H~II})=n({\rm H~II})d_{edge}$.  In this way, we estimate
$\log N({\rm H~II})=18.08\pm 0.65$, and the H ionization fraction range
implied by this value is listed in Table~3.  The total hydrogen column
(H~I+H~II) is then $\log N({\rm H})=18.47\pm 0.27$.

     We can now compute elemental abundances relative to H from the
column densities listed in Table~2.  For each element, we add the columns
of all detected ionization states.  We use the ionization state
calculations of SF02, in particular their preferred model~17, to correct
our abundances for the columns of
undetected ionization states, since we showed in \S4 that
the predictions of SF02 are consistent with our data.  For example,
model~17 suggests that O is 29.3\% ionized, so to convert our
measured O~I column density to a total O column, we add 0.15 dex to the
$\log N({\rm O~I})$ value in Table~2.  Corrections of this nature are
quite small in most cases.  The Mg~II, Al~II, Si~II, S~II, and Fe~II
columns in Table~2, for example, represent measurements of the dominant
LIC ionization states of those elements according to SF02.

     In this manner, we compute logarithmic gas phase abundances of
various elements relative to hydrogen, assuming
$\log N({\rm H})=18.47\pm 0.27$ (see above), which are listed in Table~3.
Solar photospheric abundances are also listed in the table for comparison,
as \citet{ujs01} have argued that solar abundances remain the most
plausible estimates for true LISM total abundances.  For Al and S, we
assume abundances from \citet{ng98}.  The other abundances we assume are
from more recent measurements, based on attempts to take into account
solar granulation and non-LTE (NLTE) effects in the spectral analysis.
The N, Mg, Si, and Fe abundances we assume are from \citet{hh01}, based on
the analysis of a large number of lines, with NLTE corrections.  However,
\citet{cap01,cap02} claim that the abundances of C and O can be more
accurately determined by a particularly detailed analysis of the
[C~I] $\lambda$8727 and [O~I] $\lambda$6300 forbidden lines alone, which
should be largely immune to NLTE effects.  Thus, the C and O solar
abundances listed in Table~3 are from their analyses, which use a fully
three-dimensional treatment of granulation.  The changes in solar
abundances provided by these new analyses can be significant.  The apparent
O depletion that SF02 report for the LIC disappears if the new, lower solar
O abundance is assumed.

     In many cases, the LIC gas phase abundances listed in Table~3 are
significantly below the solar abundances.  Logarithmic depletion values
can be computed by subtracting the solar abundances in Table~3 from the
LIC abundances, and in Figure~9 these depletions are plotted versus atomic
number.  The heavier elements of Mg, Al, Si, and Fe are all depleted by
factors of about $10-30$.  Presumably, the missing heavy elements have
been incorporated into dust grains.  Based on the analysis of many long
lines of sight, \citet{bds96} find depletions for warm disk material in
the Galaxy to be between $-0.73$ and $-0.90$ for Mg, $-0.35$ and $-0.51$
for Si, and $-1.19$ and $-1.24$ for Fe.  Our measurements suggest
somewhat larger depletions for the LIC towards Capella, although only the
Si value falls completely outside the ranges quoted above
considering the large error bars.

     Previous observations have generally suggested small depletions for
C, N, and O of order a factor of 2 or 3 within the LISM
\citep{tps95,ujs97,dmm97,dmm98,hwm02}.  However, these conclusions have to
be reconsidered based on the lower solar C and O abundances suggested by
\citet*{cap01,cap02}.  For the LIC, Figure~9 suggests that N may be
slightly depleted by about a factor of 2, while C and O agree well with
solar abundances, but uncertainties are of order a factor of 2 for all
these measurements.  \citet*{cap01,cap02} note that there is no N~I
forbidden line that they can analyze like [C~I] $\lambda$8727 and
[O~I] $\lambda$6300, but based on their results for C and O they speculate
that perhaps the solar N abundance from \citet{hh01} that we are using
could also be overestimated by about 0.1 dex, in which case the N abundance
in Figure~9 could be more consistent with solar than the figure
currently suggests.  Oxygen is the one element whose ionization state is
coupled to that of hydrogen to the extent that a more precise abundance
could in principle be derived simply by assuming ${\rm O/H}={\rm O~I/H~I}$,
in which case the derived depletion is $0.09\pm 0.33$.  However, this
result is not that much different from that shown in Figure~9,
computed as described above using the SF02 corrections for unobserved
ionization states.

     The upper limit of the C error bar could perhaps be lowered a little
if the S~II upper limit in Table~2 is used to further constrain the C~II
column density.  This requires additional assumptions, namely that the
ionization states of C and S are identical and that the S abundance is
close to solar.  \citet{ujs98} argue that these are reasonable assumptions
consistent with existing LISM data.  The difference in logarithmic solar
abundances of C and S is $1.06\pm 0.12$ (see Table~3).  Assuming the upper
bound of this range, the S~II upper limit in Table~2 then implies an upper
limit for C~II of $\log N({\rm C~II})<14.78$, which is well below the upper
limit of the measured $\log N({\rm C~II})=14.8\pm 0.3$ result listed in
Table~2.  This result depends on the accuracy of both the solar C and S
abundances, and the substantial revision of the solar C abundance by
\citet*{cap02} illustrates the potential danger of this assumption.
Nevertheless, we note that the revised upper limit for C~II would change
the range of electron density derived by \citet{bew97} from
$n_{e}=0.05-0.23$ cm$^{-3}$ to $n_{e}=0.11-0.23$ cm$^{-3}$, and would lower
the upper bound on the C abundance in Figure~9 by about 0.3 dex.  It would
also increase the lower bound of the C~III ionization fraction range in
Figure~8 by 0.3 dex, but this range would still be consistent with
all the SF02 models.

     Figure~10 compares our O and N abundances with measurements from
several other sources.  The square in the figure shows the total O and N
abundances measured toward G191-B2B \citep{ml02}, a white dwarf only
$7^{\circ}$ from Capella but farther away ($d=69$~pc), and containing two
ISM absorption components in addition to the LIC one.  The circle shows
the average abundances measured toward three stars within the Local Bubble,
including G191-B2B \citep{hwm02}.  Finally, the diamond shows the average
abundances measured for much longer lines of sight \citep{dmm97,dmm98}.
Our LIC measurements toward Capella are consistent with these other
measurements, and are in particularly good agreement with the G191-B2B and
Local Bubble average measurements.  The N abundance for the
longer lines of sight appears somewhat discrepant from the other
measurements, but this could actually be due to ionization state
variations rather than total N abundance variations.  Our Capella
measurements are based on measurements of both N~I and N~II, but the
others assume that ${\rm N/H}={\rm N~I/H~I}$.  This will only be true in an
approximate sense since the ionization states of N and H are not as tightly
coupled by charge exchange as is the case for O and H
\citep[see, e.g.,][]{hwm02}.  Note that we have converted all error bars
to 2$\sigma$ in Figure~10, and the \citet*{dmm98} O/H value was modified
to take into account a new absorption strength for O~I $\lambda$1356
from \citet{dew99}.

     One word of caution that we might add concerning our LIC depletion
values is that observations have shown that gas phase abundances can vary
a surprising amount over relatively short distance scales within the LISM.
Observations of $\alpha$~Cen and 36~Oph in the opposite direction from
Capella (see Fig.~1) suggest a Mg~II/D~I ratio four times higher than
toward Capella \citep*{jll96,bew00}.  Even more
remarkable is the 29~pc line of sight to $\beta$~Cet, which shows a
Mg~II/D~I ratio about 12 times higher than for Capella \citep{np97},
implying no significant Mg depletion at all.  The
$\alpha$~Cen line of sight samples mostly G cloud rather than LIC material
\citep*[assuming they are really different;][]{bew02a}, and the
$\beta$~Cet line of sight also does not exclusively sample LIC material,
but these results still suggest that abundances could vary within the LIC
and that abundances at the Sun's location could possibly be different
from the average LIC values we measure toward Capella.

\section{SUMMARY}

     We have analyzed LIC absorption lines in HST/STIS and FUSE
observations of Capella to measure column densities for as many
atomic species as possible.  Our results are summarized as follows:
\begin{description}
\item[1.] We detect D~I and H~I Ly$\beta$ absorption in the FUSE data,
  but the quality of the data do not allow us to improve on previous
  measurements of D~I and H~I from HST observations of Ly$\alpha$.
\item[2.] The combined FUSE and HST/STIS data sets allow us to measure or
  estimate low upper limits for column densities of all three of the
  lowest ionization states of C, N, and Si.  This allows us to empirically
  establish the ionization states of these elements within the LIC.  At
  least 95\% of C is in the form of C~II, at least 90\% of Si is in the
  form of Si~II, and N is roughly half-ionized.
\item[3.] The C, N, and Si ionization states are consistent with the
  predictions of steady state photoionization models for the LIC computed
  by \citet{jds02}, providing some support for the contention that
  photoionization alone can account for the observed ionization level of
  the LIC.  However, more must be known about
  the diffuse EUV background before this can truly be established.
\item[4.] Based on our column density measurements and previous ones, we
  measure total abundances for seven elements.  The heavy elements
  Mg, Al, Si, and Fe are all depleted relative to solar abundances by 
  factors of about $10-30$, presumably due to the incorporation of these
  elements into dust grains.  The abundances of carbon, nitrogen, and
  oxygen are close to solar, although the error bars are large enough
  to also allow depletions of about a factor of 2 for these elements.
  Our measurements of O/H and N/H are consistent with previous LISM
  measurements \citep{dmm97,dmm98,ml02,hwm02}.
\end{description}

\acknowledgments

We would like to thank E.\ B.\ Jenkins, G.\ H\'{e}brard, and the
referee J.\ V.\ Vallerga for useful comments on the paper.  We would
also like to thank J.\ D.\ Slavin for providing us with unpublished
results from his LIC photoionization models.
Support for this work was provided by NASA through
grants NAG5-9041 and S-56500-D to the University of Colorado.  This work
is based in part on data obtained for the Guaranteed Time Team by the
NASA-CNES-CSA FUSE mission operated by the Johns Hopkins University.
Financial support to U.\ S.\ participants has been provided by NASA
contract NAS5-32985.

\clearpage

\begin{deluxetable}{ccccccccc}
\tabletypesize{\scriptsize}
\tablecaption{Observation Summary}
\tablecolumns{9}
\tablewidth{0pt}
\tablehead{
  \colhead{Instrument} & \colhead{Data Set} & \colhead{Grating} &
    \colhead{Aperture} & \colhead{Spectral} & \colhead{Resolution} &
    \colhead{Date} & \colhead{Start Time} & \colhead{Exp.~Time} \\
  \colhead{} & \colhead{} & \colhead{} & \colhead{} & \colhead{Range (\AA)} &
    \colhead{($\lambda$/$\Delta\lambda$)} & \colhead{} & \colhead{(UT)} &
    \colhead{(s)} }
\startdata
HST/STIS&O5LC01&E140M&0.2x0.06&$1170-1710$&40,000&1999 Sept.~12&19:07&8,113\\
FUSE &P1041301&\nodata&LWRS& $905-1187$&15,000 & 2000 Nov.~5  &21:56&14,182\\
FUSE &P1041302&\nodata&LWRS& $905-1187$&15,000 & 2000 Nov.~7  & 2:22&12,332\\
FUSE &P1041303&\nodata&MDRS& $905-1187$&15,000 & 2001 Jan.~11 & 6:11&21,359\\
\enddata
\end{deluxetable}

\begin{deluxetable}{lccccc}
\tabletypesize{\small}
\tablecaption{Capella ISM Lines}
\tablecolumns{6}
\tablewidth{0pt}
\tablehead{
  \colhead{Species} & \colhead{$\lambda_{rest}$(\AA)} &
    \colhead{Instrument\tablenotemark{a}} & \colhead{f} &
    \colhead{$\log N$\tablenotemark{b}} & \colhead{Ref.} }
\startdata
H I    & 1215.670 & HST/GHRS     & $4.16\times 10^{-1}$ & $18.24\pm 0.07$ & 1,2 \\
D I    & 1215.339 & HST/GHRS     & $4.16\times 10^{-1}$ & $13.444\pm 0.016$ & 1 \\
C I    & 1656.928 & HST/STIS     & $1.41\times 10^{-1}$ & $<12.5$           & 3 \\
C II   & 1334.532 & HST/GHRS     & $1.28\times 10^{-1}$ & $14.8\pm 0.3$     & 4 \\
C II$^{*}$&1335.708&HST/GHRS     & $1.15\times 10^{-1}$ & $12.64\pm 0.07$   & 4 \\
C III  &  977.020 & FUSE (SiC2A) & $7.62\times 10^{-1}$ & $13.02\pm 0.13$   & 3 \\
C IV   & 1548.195 & HST/STIS     & $1.91\times 10^{-1}$ & $<11.9$           & 3 \\
N I    & 1199.550 & HST/STIS     & $1.33\times 10^{-1}$&$13.86\pm 0.17$\tablenotemark{c}& 3\\
       & 1200.223 & HST/STIS     & $8.85\times 10^{-2}$&$13.86\pm 0.17$\tablenotemark{c}& 3\\
       & 1200.710 & HST/STIS     & $4.42\times 10^{-2}$&$13.86\pm 0.17$\tablenotemark{c}& 3\\
N II   & 1083.990 & FUSE (SiC2B) & $1.03\times 10^{-1}$ & $13.62\pm 0.46$   & 3 \\
N III  &  989.799 & FUSE (SiC2A) & $1.07\times 10^{-1}$ & $<12.9$           & 3 \\
O I    &  988.655 & FUSE (SiC2A) & $7.71\times 10^{-3}$&$14.93\pm 0.32$\tablenotemark{d}& 3\\
       &  988.773 & FUSE (SiC2A) & $4.32\times 10^{-2}$&$14.93\pm 0.32$\tablenotemark{d}& 3\\
       & 1039.230 & FUSE (LiF1A) & $9.20\times 10^{-3}$&$15.01\pm 0.44$\tablenotemark{d}& 3\\
       & 1302.169 & HST/STIS     & $4.89\times 10^{-2}$&$15.14\pm 0.35$\tablenotemark{d}& 3\\
Mg II  & 2796.352 & HST/GHRS     & $6.12\times 10^{-1}$ & $12.809\pm 0.022$ & 1 \\
       & 2803.531 & HST/GHRS     & $3.05\times 10^{-1}$ & $12.809\pm 0.022$ & 1 \\
Al II  & 1670.787 & HST/STIS     & $1.83\times 10^{0}$  & $11.43\pm 0.08$   & 3 \\
Si I   & 1562.002 & HST/STIS     & $3.76\times 10^{-1}$ & $<11.9$           & 3 \\
Si II  & 1260.422 & HST/STIS     & $1.01\times 10^{0}$ &$12.97\pm 0.08$\tablenotemark{c}& 3\\
       & 1304.370 & HST/STIS     & $1.47\times 10^{-1}$&$12.97\pm 0.08$\tablenotemark{c}& 3\\
       & 1526.707 & HST/STIS     & $2.30\times 10^{-1}$&$12.97\pm 0.08$\tablenotemark{c}& 3\\
Si III & 1206.500 & HST/STIS     & $1.67\times 10^{0}$  & $<10.9$           & 3 \\
S II   & 1259.519 & HST/STIS     & $1.62\times 10^{-2}$ & $<13.6$           & 3 \\
Ar I   & 1048.220 & FUSE (LiF1A) & $2.44\times 10^{-1}$ & $<13.5$           & 3 \\
Fe II  & 2600.173 & HST/GHRS     & $2.24\times 10^{-1}$ & $12.494\pm 0.023$ & 1 \\
\enddata
\tablenotetext{a}{The instrument used to observe the line, and for FUSE
  the individual spectral segment used (in parentheses).}
\tablenotetext{b}{The quoted uncertainties should be considered to be
  2$\sigma$ error bars (see text).}
\tablenotetext{c}{The 3 N~I lines were fitted simultaneously, as were the
  3 Si~II lines.}
\tablenotetext{d}{Based on these separate O~I measurements, we quote a
  best estimate for the O~I column density of
  $\log N({\rm O~I})=15.02\pm 0.32$.}
\tablerefs{(1) Linsky et al.\ 1995. (2) Vidal-Madjar \& Ferlet 2002. (3)
  This paper. (4) Wood \& Linsky 1997.}
\end{deluxetable}

\begin{deluxetable}{cccccccc}
\tabletypesize{\normalsize}
\tablecaption{Element Abundances and Ionization States for the LIC Toward
  Capella}
\tablecolumns{8}
\tablewidth{0pt}
\tablehead{
  \colhead{Element} & & \multicolumn{2}{c}{Log.\ Abundance Rel.\ to H} & &
    \multicolumn{3}{c}{Ionization Fraction} \\
  \colhead{} & \colhead{} & \colhead{LIC} &
    \colhead{Solar\tablenotemark{a}} & \colhead{} &
    \colhead{I} & \colhead{II} & \colhead{III} }
\startdata
H & & 0              &  0               & &$0.323-0.814$ & $0.186-0.677$ &
 \nodata \\
C & &$-3.66\pm 0.40$ & $-3.61\pm 0.04$  & &$0-0.015$ & $0.948-0.994$     &
 $0.006-0.043$ \\
N & &$-4.41\pm 0.34$ & $-4.069\pm 0.111$& &$0.276-0.881$ & $0.112-0.711$ &
 $0-0.114$ \\
O & &$-3.30\pm 0.42$ & $-3.31\pm 0.05$  & &\nodata       & \nodata       &
 \nodata \\
Mg& &$-5.58\pm 0.27$ & $-4.462\pm 0.060$& &\nodata       & \nodata       &
 \nodata \\
Al& &$-7.02\pm 0.28$ & $-5.53\pm 0.07$  & &\nodata       & \nodata       &
 \nodata \\
Si& &$-5.50\pm 0.28$ & $-4.464\pm 0.049$& &$0-0.093$     & $0.899-1$     &
 $0-0.010$ \\
S & &$<-4.58$        & $-4.67\pm 0.11  $& &\nodata       & \nodata       &
 \nodata \\
Fe& &$-5.96\pm 0.27$ & $-4.552\pm 0.082$& &\nodata       & \nodata       &
 \nodata \\
\enddata
\tablenotetext{a}{From Grevesse \& Sauval (1998) [Al, S]; Holweger (2001)
  [N, Mg, Si, Fe]; and Allende Prieto et al.\ (2001, 2002) [C, O].}
\end{deluxetable}

\clearpage

\begin{figure}
\plotfiddle{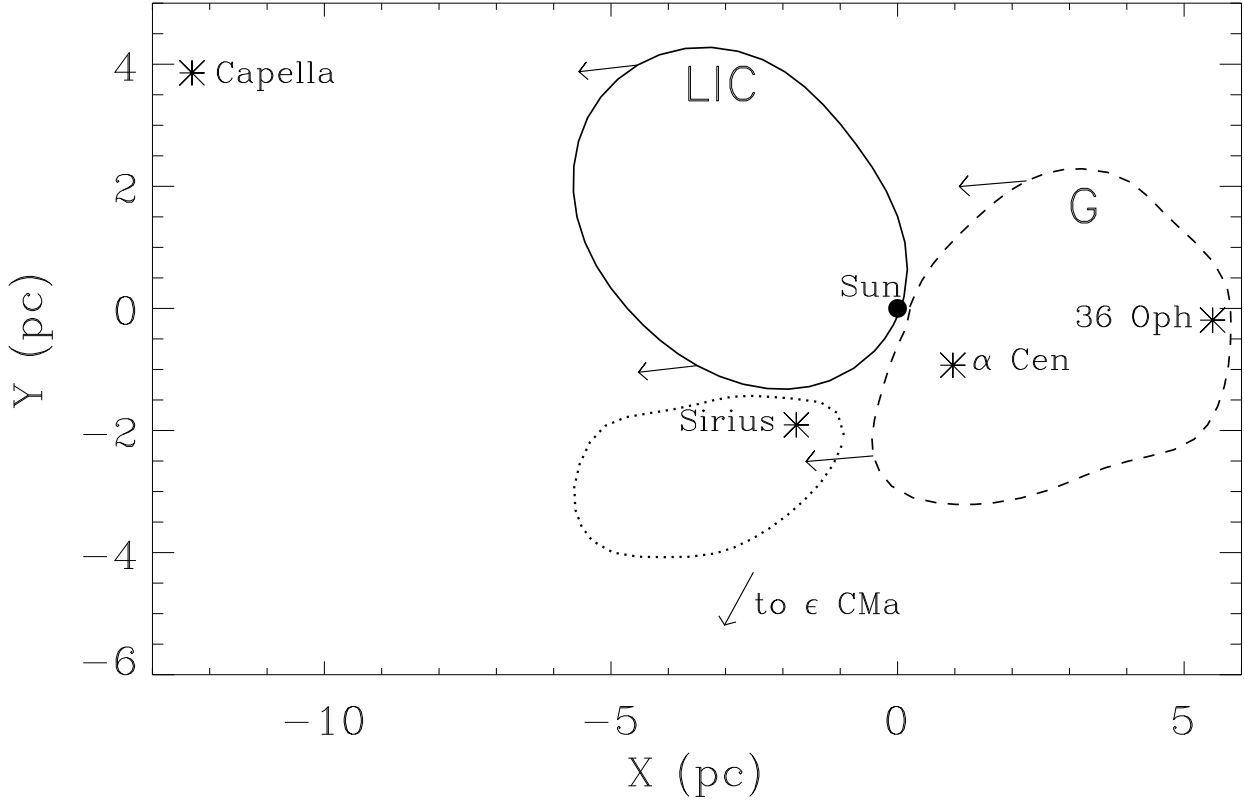}{3.5in}{90}{75}{75}{280}{0}
\caption{A map showing the locations of several stars, the LIC (solid
  line), G cloud (dashed line), and a third cloud observed towards Sirius
  and $\epsilon$~CMa (dotted line), projected onto the Galactic plane, where
  Galactic Center is to the right.  The LIC shape is
  from the Redfield \& Linsky (2000) model, while the shapes of the other
  two clouds are just rough estimates.  Arrows indicate the velocity vectors
  of the LIC and G cloud relative to the Sun.}
\end{figure}

\clearpage

\begin{figure}
\plotfiddle{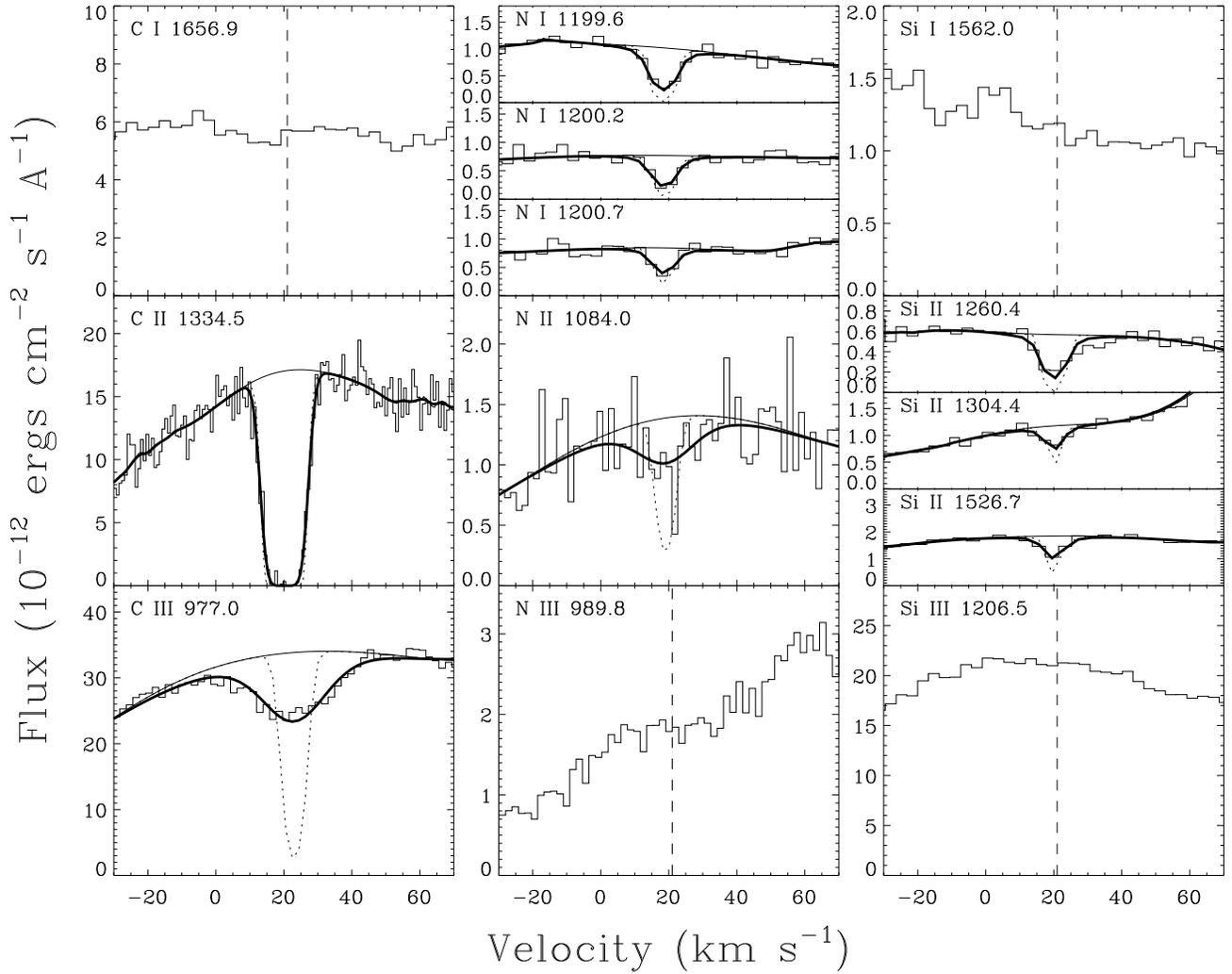}{3.5in}{90}{75}{75}{310}{10}
\caption{Detected and undetected LIC absorption lines of the lowest three
  ionization states of C (left column), N (middle column), and Si (right
  column) from HST and FUSE observations of Capella, plotted on a
  heliocentric velocity scale.  Vertical dashed lines
  mark the expected locations of undetected LIC lines.  Fits are shown to
  detected lines, where dotted and thick solid lines are the fits before
  and after convolution with the instrumental profile, respectively.}
\end{figure}

\clearpage

\begin{figure}
\plotfiddle{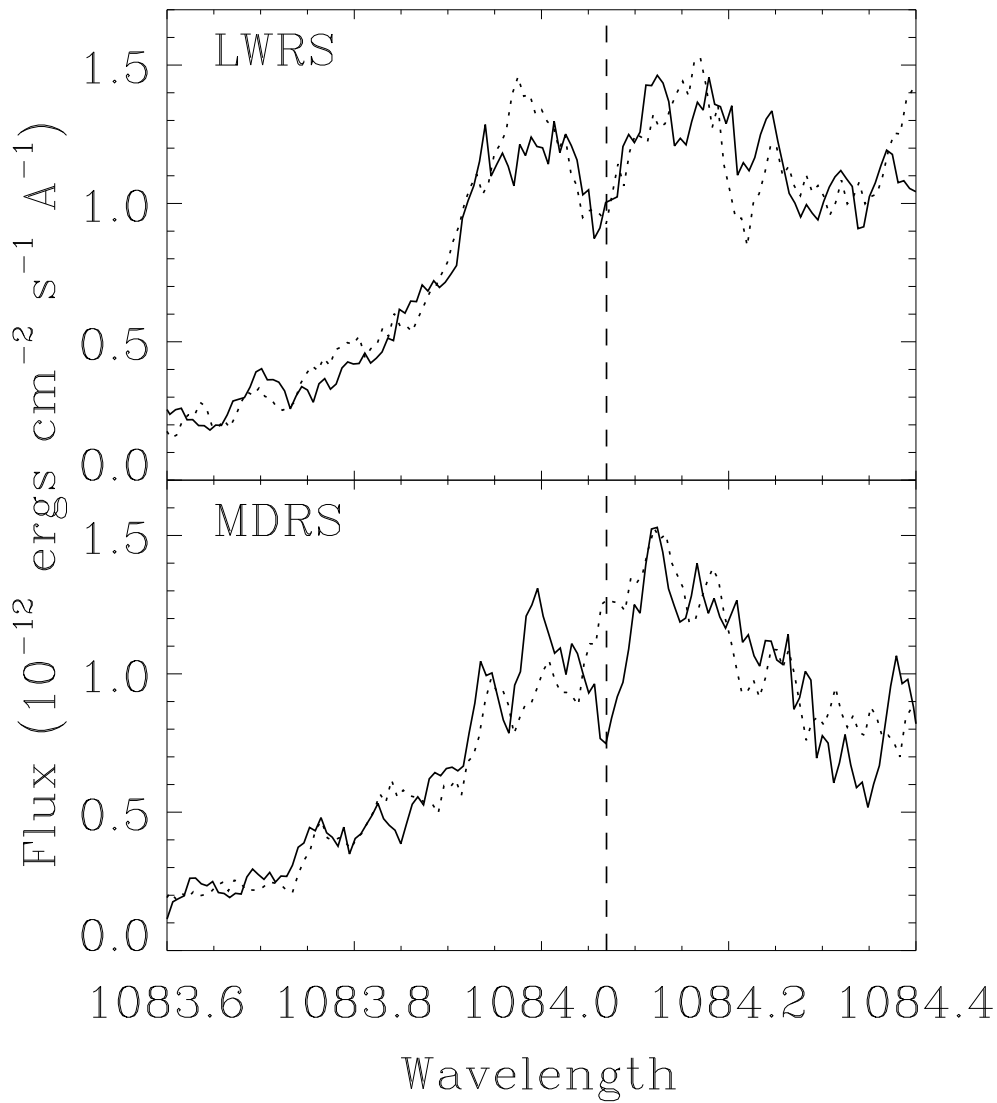}{5.5in}{0}{80}{80}{-270}{20}
\caption{Slightly smoothed FUSE LWRS and MDRS spectra of the
  N~II $\lambda$1084.0 line, where the solid lines are the SiC2B segments
  and the dotted lines are SiC1A.  The dashed line marks the location of
  the LIC absorption.  Note that it is not seen in the MDRS SiC1A data.}
\end{figure}

\clearpage

\begin{figure}
\plotfiddle{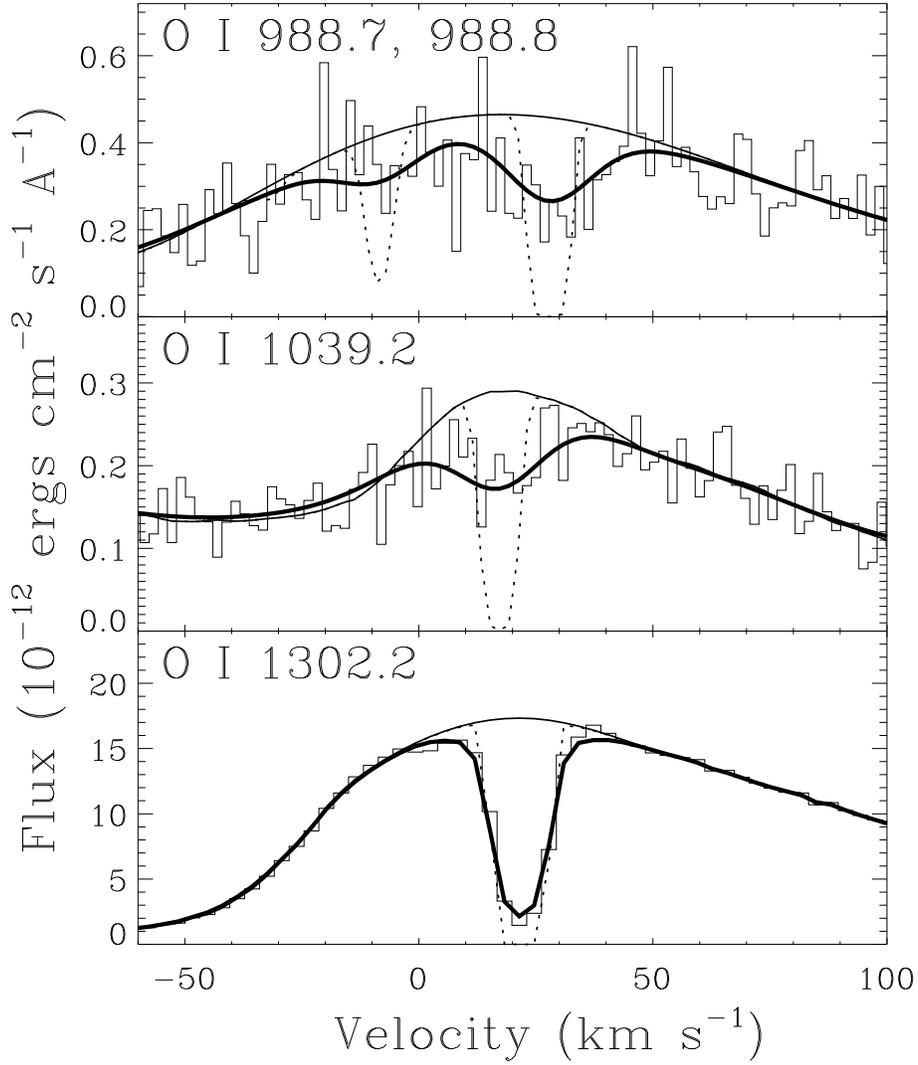}{5.5in}{0}{80}{80}{-270}{5}
\caption{Fits to LIC O~I absorption lines observed toward Capella by
  HST/STIS and FUSE, plotted on a heliocentric velocity scale.  Dotted and
  thick solid lines indicate the fits before and after convolution with
  the instrumental profile, respectively.}
\end{figure}

\clearpage

\begin{figure}
\plotfiddle{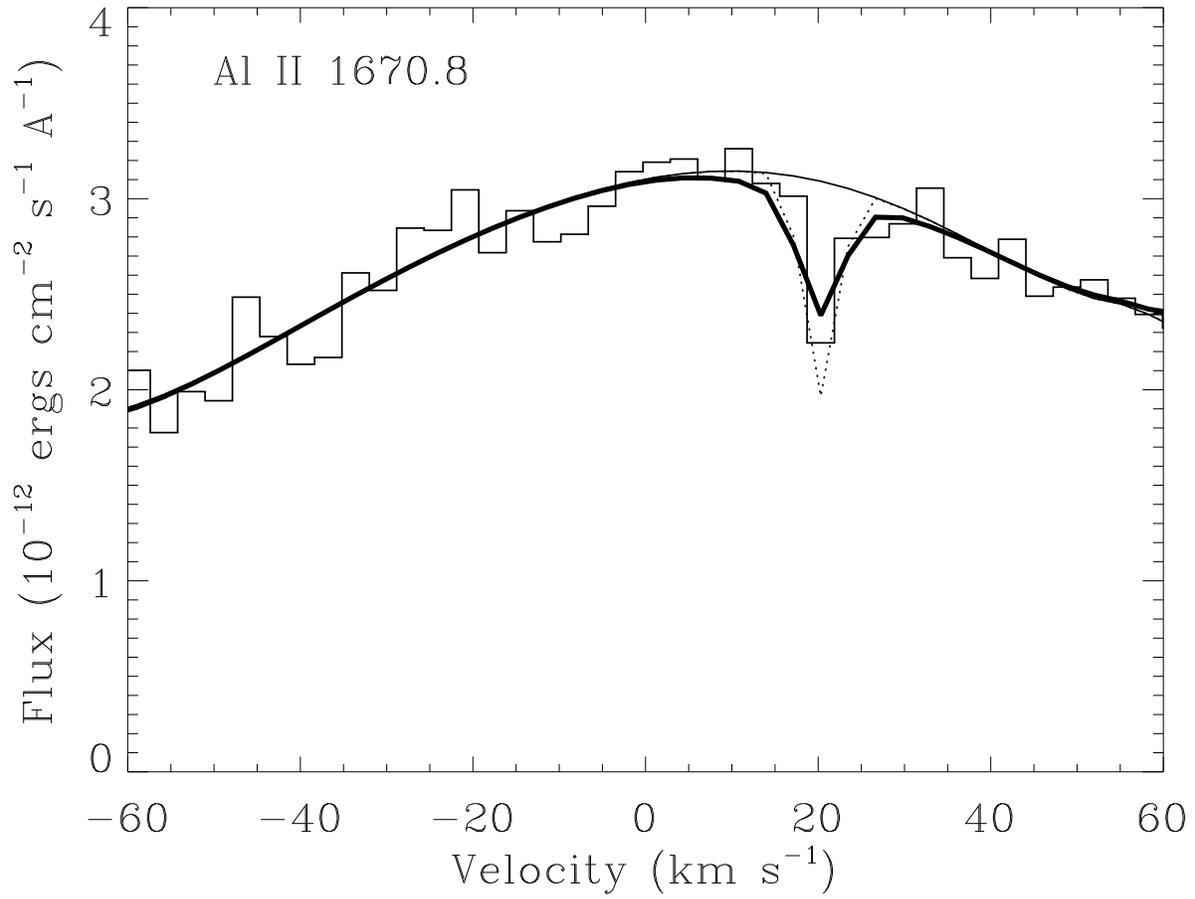}{3.5in}{90}{75}{75}{280}{0}
\caption{A fit to the Al~II $\lambda$1670.8 LIC absorption line observed
  in HST/STIS observations of Capella, where the dotted and thick solid
  lines are before and after convolution with the instrumental profile,
  respectively.}
\end{figure}

\clearpage

\begin{figure}
\plotfiddle{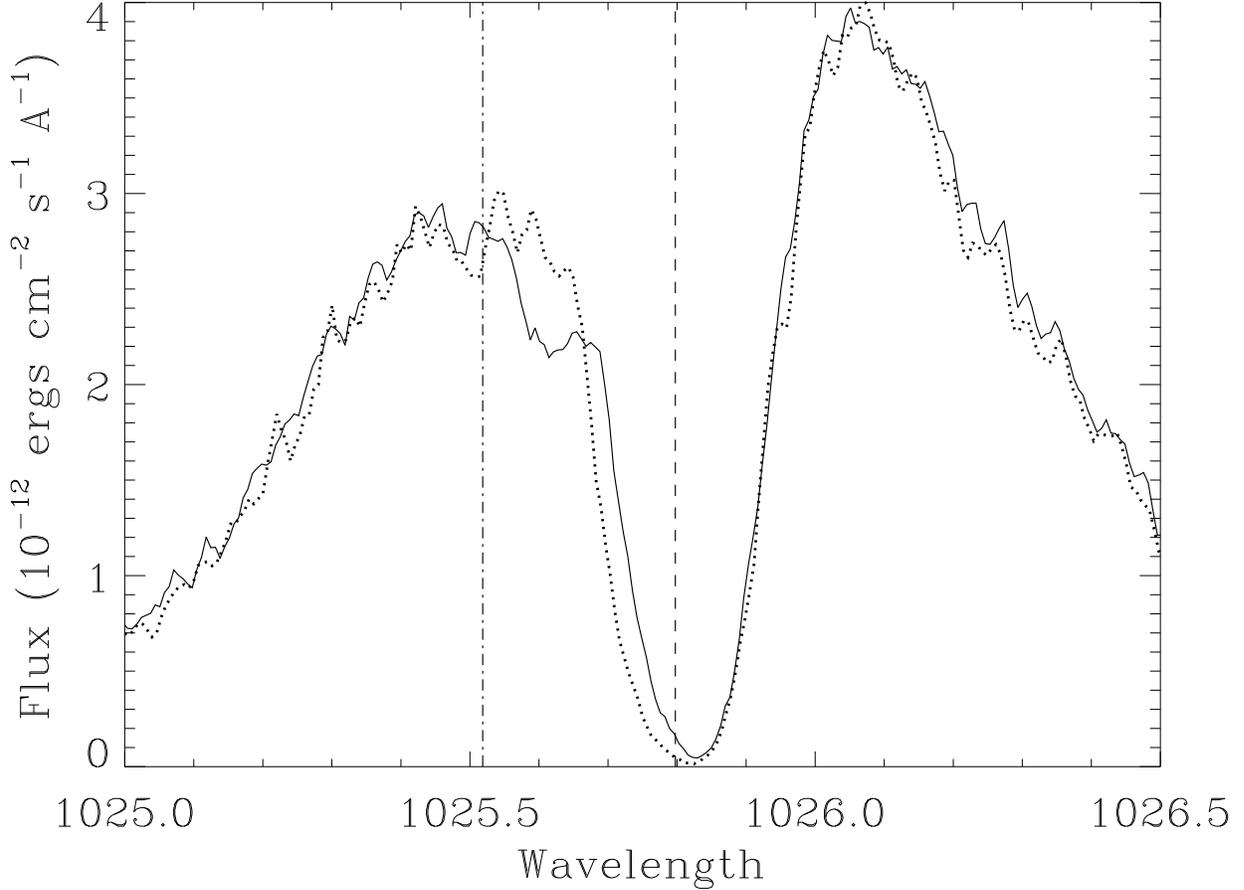}{3.5in}{90}{75}{75}{280}{0}
\caption{Two versions of the FUSE/MDRS LiF1A spectrum of Capella's Ly$\beta$
  line, one processed using CALFUSE 1.8.7 (solid line), and one processed
  using CALFUSE 2.0.5 (dotted line).  The vertical dashed and dot-dashed
  lines indicate the expected locations of LIC H~I and D~I absorption,
  respectively.  The weak D~I feature for the CALFUSE 1.8.7 spectrum
  is significantly redshifted from its expected location due to the
  uncorrected ``walk'' effect (see text), while the situation is improved
  for the CALFUSE 2.0.5 data.}
\end{figure}

\clearpage

\begin{figure}
\plotfiddle{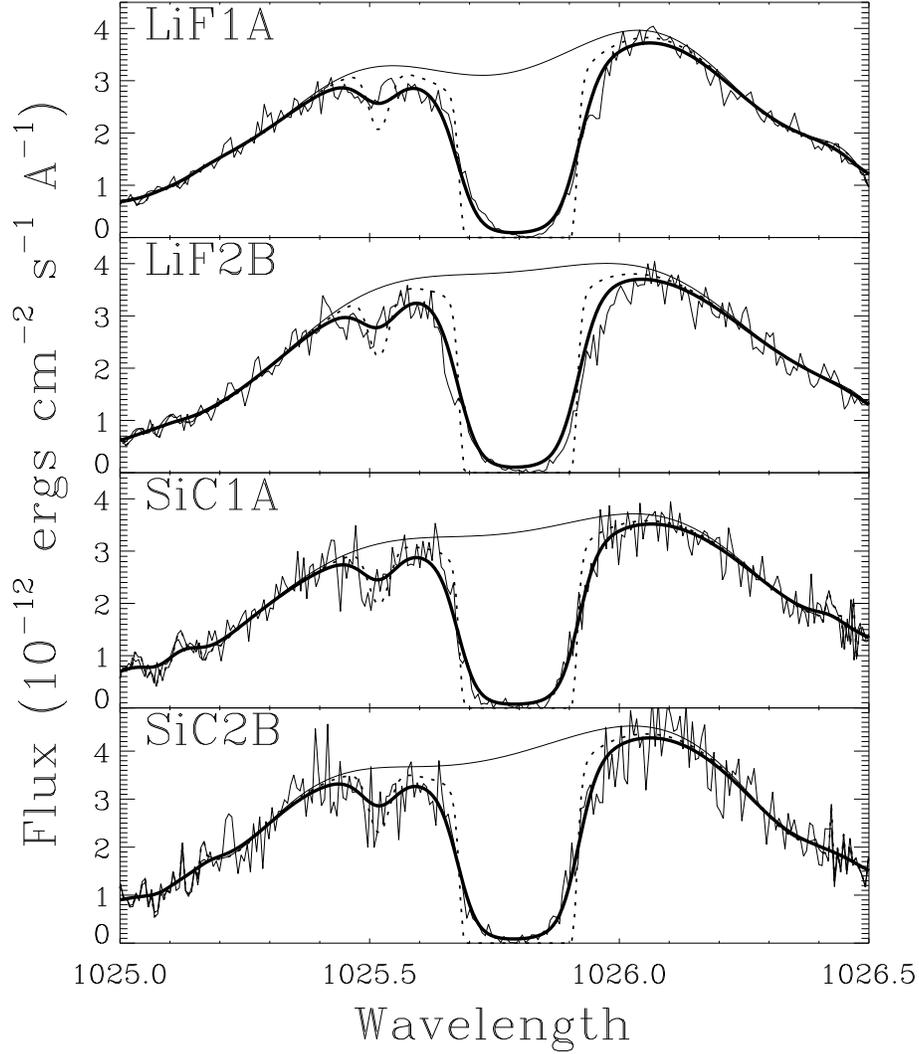}{5.5in}{0}{80}{80}{-270}{0}
\caption{Four FUSE/MDRS segments containing the Ly$\beta$ line of Capella,
  all processed with CALFUSE 2.0.5.  Estimates of the intrinsic stellar
  line profile are derived for each spectrum (thin solid lines).  The
  thick solid lines show predicted LIC H~I and D~I absorption profiles
  based on H~I and D~I column densities previously measured by Linsky
  et al.\ (1995) from HST observations of H~I and D~I Ly$\alpha$.  The
  dotted lines are these profiles before convolution with the instrumental
  profile.}
\end{figure}

\clearpage

\begin{figure}
\plotfiddle{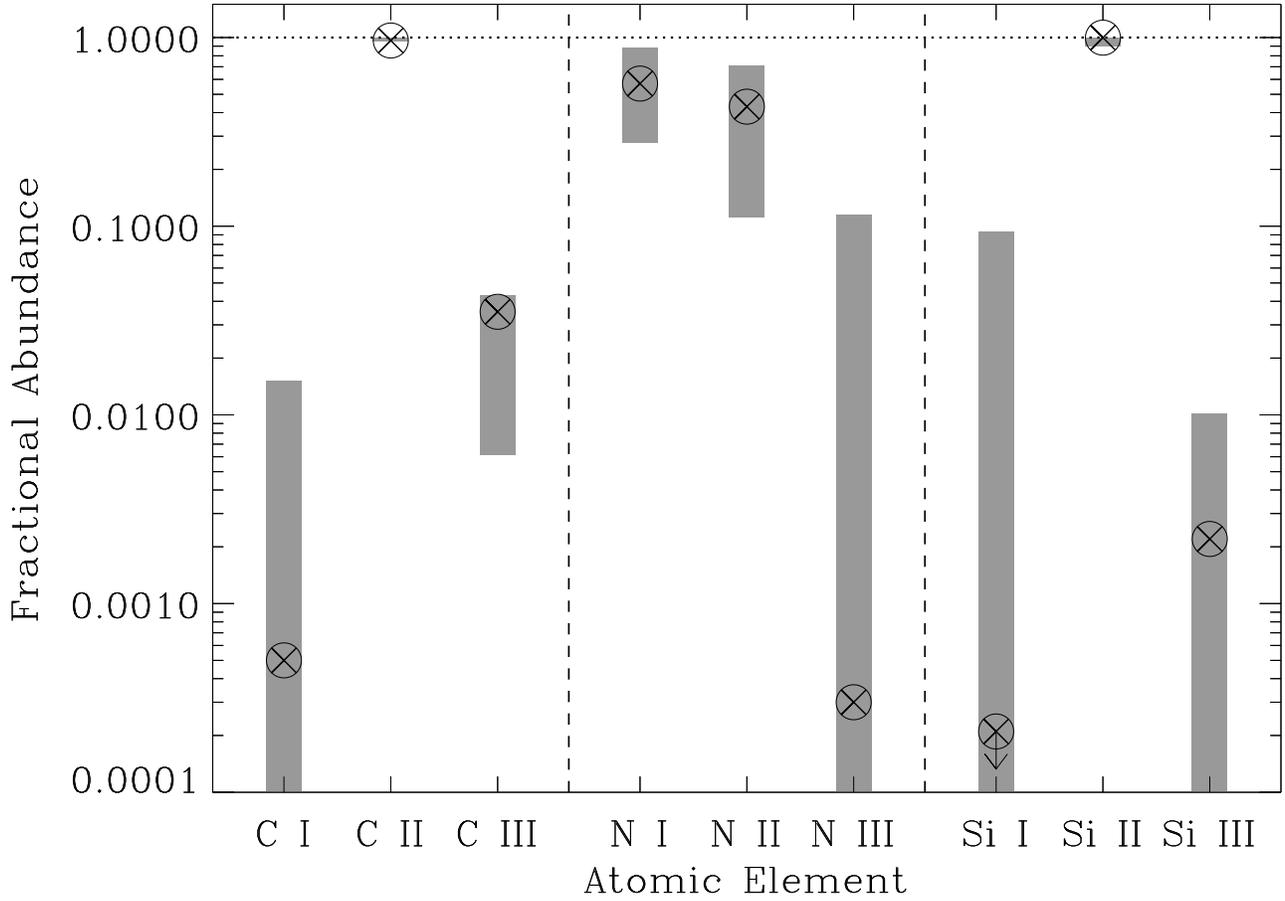}{3.5in}{90}{75}{75}{300}{0}
\caption{Fractional LIC abundances toward Capella for the three lowest
  ionization states of carbon, nitrogen, and silicon.  The shaded regions
  are the ranges allowed by column density measurements from HST and FUSE
  data, while the crossed circles indicate the predicted abundances
  of a steady state photoionization model of the LIC (model 17) from
  Slavin \& Frisch (2002).  There is good agreement between this model
  and the data.}
\end{figure}

\clearpage

\begin{figure}
\plotfiddle{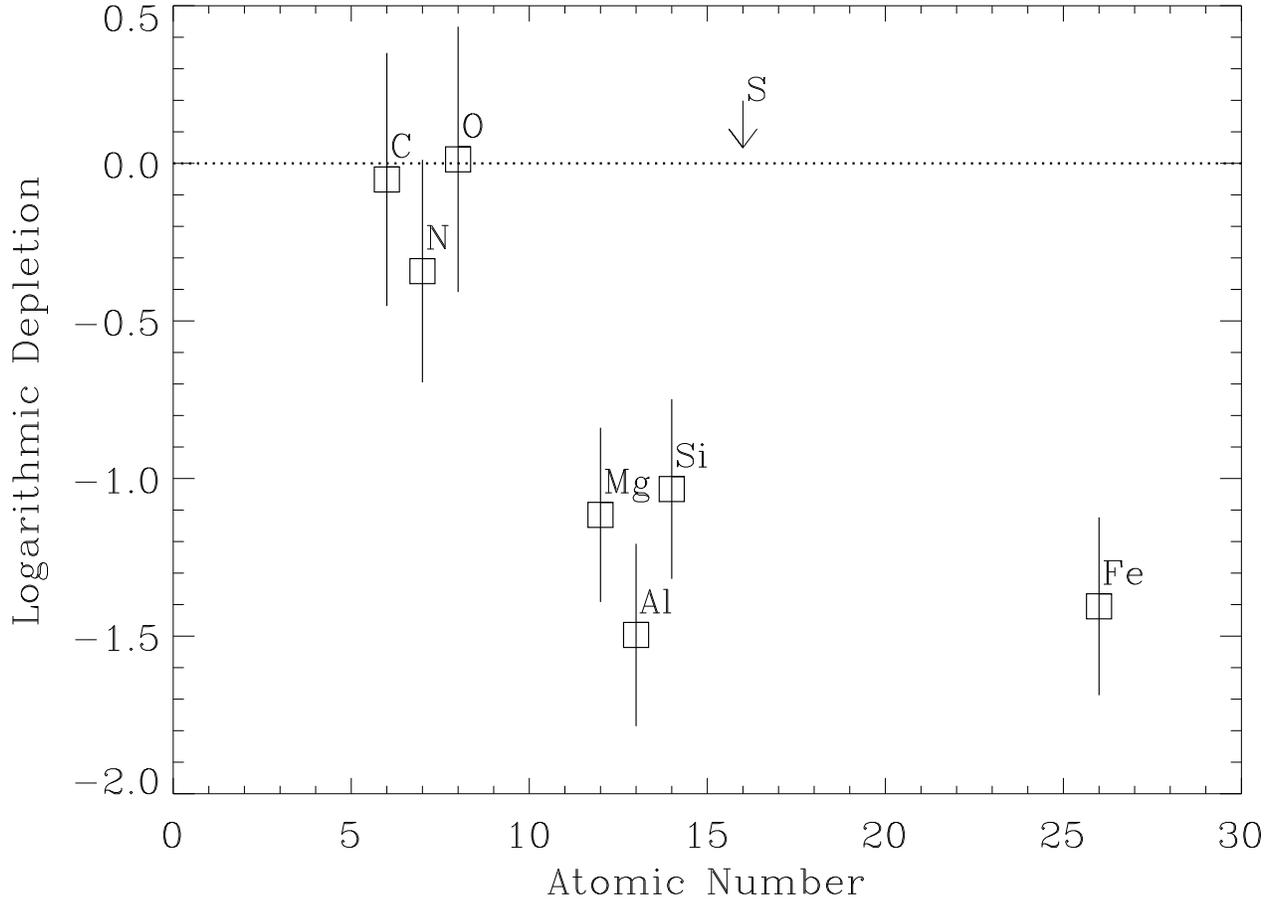}{3.5in}{90}{75}{75}{290}{0}
\caption{Logarithmic depletions of various elements for the LIC toward
  Capella.  Note that S is an upper limit.  The error bars are discussed in
  the text, and the reference solar abundances used in deriving the
  depletions are listed in Table~3.}
\end{figure}

\begin{figure}
\plotfiddle{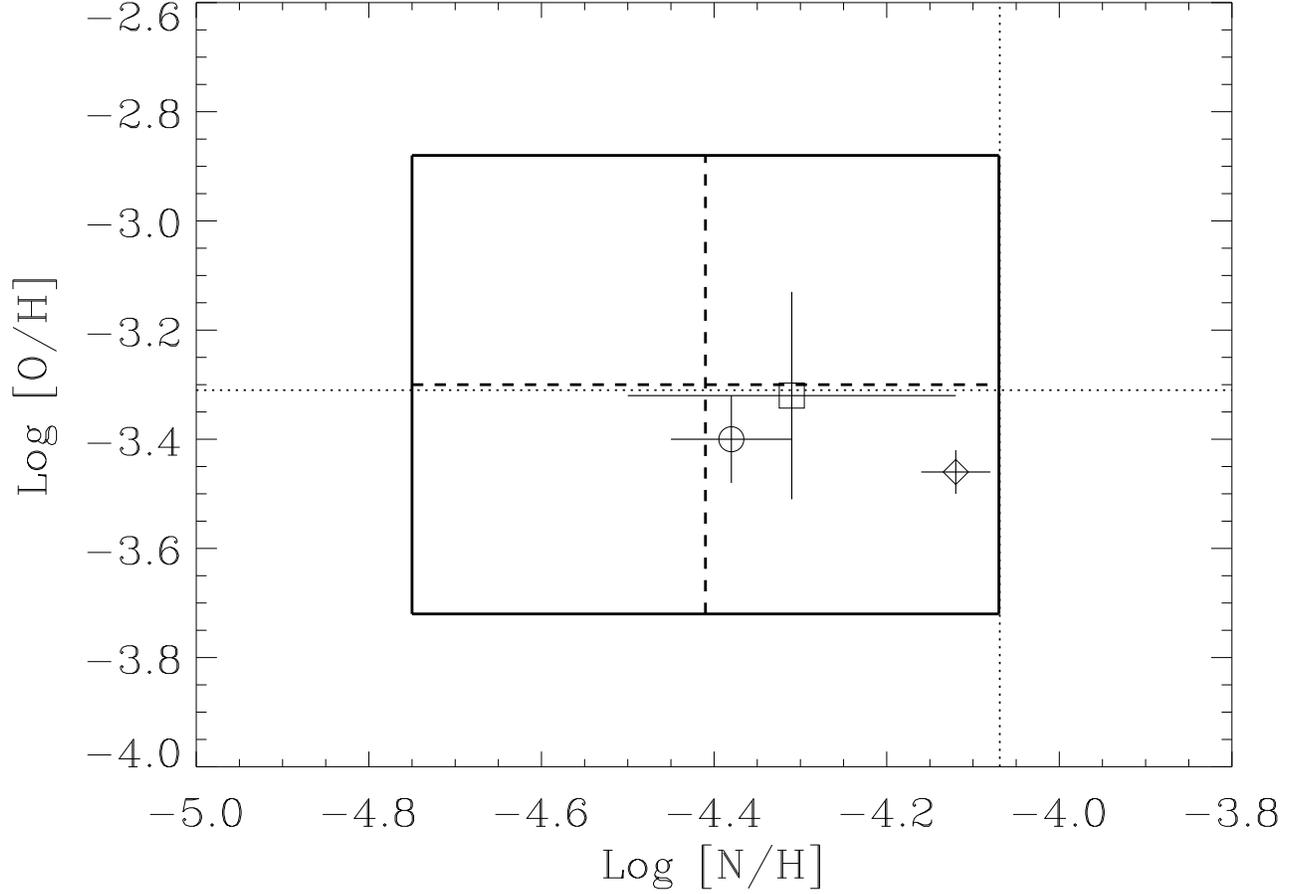}{3.5in}{90}{75}{75}{290}{0}
\caption{Oxygen versus nitrogen abundances, where the large solid rectangle
  is the error box for our measurements toward Capella, and the dotted
  lines indicate solar abundances.  The other data points plotted are for
  G191-B2B (square) from Lemoine et al.\ (2002), an average of three Local
  Bubble lines of sight (circle) from Moos et al.\ (2002), and an average
  of many longer lines of sight (diamond) from Meyer et al.\ (1997, 1998).}
\end{figure}

\end{document}